
{
\input phyzzx
\NPrefs
\def\define#1#2\par{\def#1{\Ref#1{#2}\edef#1{\noexpand\refmark{#1}}}}
\def\con#1#2\noc{\let\?=\Ref\let\<=\refmark\let\Ref=\REFS
         \let\refmark=\undefined#1\let\Ref=\REFSCON#2
         \let\Ref=\?\let\refmark=\<\refsend}

\define\DDK
F.~David, Mod. Phys. Lett. {\bf A3} (1988) 1651;
J.~Distler and H.~Kawai, Nucl. Phys. {\bf B321} (1989) 509.

\define\POLYAKOV
A.~M.~Polyakov, Phys. Lett. {\bf 103B} (1981) 207.

\define\MATRIXONE
E.~Brezin and V.~Kazakov, Phys. Lett. {\bf B236} (1990) 144;
M.~Douglas and S.~Shenker, Nucl.~Phys. {\bf B335} (1990) 635;
D.~Gross and A.~Migdal, Phys. Rev. Lett. {\bf 64} (1990) 127.

\define\MATRIXTWO
E.~Brezin, V.~Kazakov and A.~Zamolodchikov, Nucl.~Phys. {\bf B338}
(1990) 673;
D.~Gross and N.~Miljkovic, Phys.~Lett. {\bf B238} (1990) 217;
P.~Ginsparg and J.~Zinn-Justin, Phys.~Lett. {\bf B240} (1990) 333;
S.~Das, A.~Dhar, A.~M.~Sengupta and S.~Wadia, Mod.~Phys.~Lett. {\bf A5}
(1990) 891;
G.~Parisi, Phys.~Lett. {bf B238} (1990) 209;
D.~Gross and A.~Klebanov, Nucl.~Phys. {\bf B334} (1990) 475.

\define\GROSSMIGDAL
D.~Gross and A.~A.~Migdal, Nucl.~Phys. {\bf B340} (1990) 333.

\define\TOPOLOGY
E.~Witten, Nucl.~Phys. {\bf B340} (1990) 280;
R.~Dijkgraaf and E.~Witten, Nucl.~Phys. {\bf B342} (1990) 486;
J.~Distler, Nucl.~Phys. {\bf B342} (1990) 523;
E.~Verlinde and H.~Verlinde, Nucl. Phys. {\bf 348} (1991) 457;
R.~Dijkgraaf, E.~Verlinde and H.~Verlinde, Nucl.~Phys. {\bf B348}
(1991) 435.

\define\GOULIANLI
M.~Goulian and M.~Li, Phys.~Rev.~Lett. {\bf 66} (1991) 2051;
P.~Di Francesco and D.~Kutasov, Phys.~Lett. {\bf B261} (1991) 261 and
PUPT-1276, (September, 1991);
Y.~Kitazawa, Phys.~Lett. {\bf B265} (1991) 262;
K.~Aoki and E.~D'Hoker, UCLA Preprint, UCLA/91/TEP/32.

\define\DOTSENKO
V.~Dotsenko, Paris Preprint PAR-LPTHE 91-18, (February, 1991);
Carg\` ese lecture, PAR-LPTHE 91-52, (October, 1991).

\define\THORNETAL
T.~Curtright and C.~Thorn, Phys.~Rev.~Lett. {\bf 48} (1982) 1309;
E.~Braaten, T.~Curtright and C.~Thorn, Phys.~Lett. {\bf B118}
(1982) 115,
Ann.~Phys. {\bf 147} (1983) 365;
E.~Braaten, T.~Curtright, G.~Ghandour and C.~Thorn, Phys.~Rev.~Lett.
{\bf 51} (1983) 19;
E.~D'Hoker and R.~Jackiw, Phys.~Rev. {\bf D26} (1982) 3517;
J.-L.~Gervais and A.~Neveu, Nucl.~Phys. {\bf B209} (1982) 125,
{\bf B224} (1983) 329.

\define\GUPTA
A.~Gupta, S.~Trivedi and M.~Wise, Nucl. Phys. {\bf B340} (1990) 475.

\define\KPZ
A.~M.~Polyakov, Mod.~Phys.~Lett. {\bf A2} (1987) 893;
V.~G.~Knizhnik, A.~M.~Polyakov and A.~B.~Zamolodchikov, Mod. Phys.
Lett. {\bf A3} (1988) 819.

\define\KAZAMIG
V.~Kazakov and A.~Migdal, Nucl.~Phys. {\bf B311} (1989) 171;
I.~Kostov, Phys.~Lett. {\bf B215} (1988) 499;
D.~Gross, I.~Klebanov and M.~Newman, Nucl.~Phys. {\bf B350} (1990) 621.

\define\DIFRKUTMATRIX
P.~Di Francesco and D.~Kutasov, Nucl.~Phys. {\bf B342} (1990)589 and
Carg\` ese lecture, PUPT-1206 (1990).

\define\IMBIMUKHI
C.~Imbimbo and S.~Mukhi, Nucl.~Phys. {\bf B364} (1991) 662.

\define\PRIVATE
C.~Imbimbo and S.~Mukhi, private communication.

\define\DIFRANSALZ
P.~Di Francesco, H.~Saleur and J.-B.~Zuber, J.~Stat.~Phys. {\bf 49}
(1987) 57; V.~Pasquier, J.~Phys. {\bf A 20} (1987) L1229.

\define\ZEMICLASSICAL
A.~B.~Zamolodchikov, Phys. Lett. {\bf B117} (1982) 87.

\define\LIANZ
B.~Lian and G.~Zuckerman, Phys.~Lett. {\bf B254} (1991) 417, {\bf B266}
(1991) 21.

\define\TIFRBRST
S.~Mukherji, S.~Mukhi and A.~Sen, Phys.~Lett. {\bf B266} (1991) 337;
C.~Imbimbo, S.~Mahapatra and S.~Mukhi, Genova and Tata Preprint
GEF-TH-8/91, TIFR-TH-91-27 (to appear in Nucl.~Phys. B);
S.~Mukhi, Carg\` ese lecture, TIFR-TH-91-50.

\define\BMPITOHTA
P.~Bouwktnegt, J.~McCarthy and K.~Pilch, CERN Preprint CERN-TH-6162/91
(July, 1991);
K.~Itoh and N.~Ohta, Fermilab, Osaka and Brown Univ. Preprint,
FERMILAB-PUB-91/228-T, (September, 1991).

\define\BERKLEV
M. Bershadsky and I. Klebanov, Phys.~Rev.~Lett {\bf 65} (1990) 3088;
Nucl.~Phys. {\bf B360} (1991) 559;
U.~Danielsson, Princeton Preprint, PUPT-1199, (September, 1990).

\define\DOTSENKOFATEEV
V.S. Dotsenko and V.A. Fateev, Nucl. Phys. {\bf B240} (1984) 312;
{\bf B251} (1985) 691.

\define\SEIBERG
N.~Seiberg, Rutgers Preprint, RU-90-29;
L.~Alvarez-Gaum\' e and C.~G\' omez, CERN Preprint, CERN-TH-6175/91;
E.~D'Hoker, UCLA Preprint, UCLA/91/TEP/41.

\define\SAKAITANII
N.~Sakai and Y.~Tanii, Tokyo Preprint, TIT/HEP-168, (March, 1991).

\define\FELDER
G.~Felder, Nucl.~Phys. {\bf B317} (1989) 215.

\def\cc{{\cal C}}
\def\cd{{\cal D}}
\def\half{{1\over 2}}
\def\co{{\cal O}}
\def\d{\delta}
\def\dnot{\delta_0}

\def\l{\langle}
\def\r{\rangle}
\def\lalrs{\alpha_{r,s}^L}
\def\malrs{\alpha_{r,s}^M}
\def\laln{\alpha_{n',n}^{L-}}
\def\lals{\alpha_{s',s}^{L-}}
\def\lalm{\alpha_{m',m}^{L-}}
\def\lb{\beta_L}
\def\mb{\beta_M}
\def\lphi{\varphi^L}
\def\mphi{\varphi^M}
\def\ql{Q_L}
\def\qm{Q_M}
\def\id{\hbox{\bf 1}}
\def\g#1{\Gamma(#1)}
\pubnum{91-58}\date{December, 1991}
\titlepage

\title{THREE-POINT FUNCTIONS OF NON-UNITARY MINIMAL MATTER
COUPLED TO GRAVITY}
\author{Debashis Ghoshal and Swapna Mahapatra}\foot{e-mail addresses:
GHOSHAL@TIFRVAX.BITNET and SWAPNA@TIFRVAX.BITNET}
\address{Tata Institute of Fundamental Research, Homi Bhabha Road,
Bombay 400 005, India}
\bigskip
\abstract
The tree-level three-point correlation functions of local operators
in the general
$(p,q)$ minimal models coupled to gravity are calculated in the
continuum
approach. On one hand, the result agrees with the unitary series
($q=p+1$); and on the other hand, for $p=2, q=2k-1$,
we find agreement with the one-matrix model results.
\endpage
A non-perturbative understanding of gravity in two (and less than two)
dimensions has been made possible in recent years by the matrix model
approach \MATRIXONE\MATRIXTWO\ and through the developments in two
dimensional
topological gravity \TOPOLOGY. While some concrete progress have been
made, the more traditional continuum approach largely remains
ill-understood. It is therefore very important to understand various
aspects of this Liouville theory \POLYAKOV\ in the context of
non-critical
string theory and quantum gravity \KPZ\DDK. Inspite of the
complications of the Liouville dynamics
(for review see \SEIBERG\ and references therein), the so called
free-field approach has turned out to be very successful. This has been
strengthened in recent times by the zero-mode integration technique
\GUPTA\
and the subsequent calculation of the three-point function
\GOULIANLI; the calculation of three-point
functions by adding marginal perturbations to the free Liouville action
\DOTSENKO; the computation of the torus partition function
\BERKLEV\ and through the BRST analysis of the physical states of
minimal matter coupled to gravity \LIANZ\TIFRBRST\BMPITOHTA.

The computation of the three-point functions in the continuum approach
has
so far been restricted to the case of unitary minimal models coupled to
gravity. In ref.\GOULIANLI\ Goulian and Li used a Coulomb gas type
formalism with the cosmological term as the screening charge in the
Liouville theory and they used analytic continuation to deal with the
fractional power of this term. Subsequently Dotsenko \DOTSENKO\ proposed
a modification of the free Liouville action by a marginal
{\it perturbation} by
the cosmological term and a {\it conjugate} cosmological term, which
serve
as the two types of screening charges \DOTSENKOFATEEV. The coupling of
the conjugate cosmological operator is tuned in a very particular
way and
it is related to the cosmological constant $\mu$. The advantage of this
approach is that the screening charges are raised to an integral (albeit
negative at times) power  and the analytic continuation procedure is
simpler to carry out.

In this letter, we compute the tree-level three-point functions of the
general non-unitary $(p,q)$
minimal models coupled to gravity by perturbing the free Liouville
action
by two mutually conjugate screening operators, and tuning their
corresponding couplings to appropriate values that are related to the
cosmological constant $\mu$. (The perturbations we consider are by the
gravitationally dressed matter identity and they are not
the cosmological operators for the non-unitary theories.)
For $q=p+1$, the minimal matter is
unitary and our action coincides with that of ref.\DOTSENKO. For other
values of $q$, the minimal matter is non-unitary and we derive a general
formula for the three-point function. We compare this result, for the
particular case of $p=2$, with the one-matrix model results of
ref.\GROSSMIGDAL\ and show that they agree. We also comment on the
difficulties involved in working with the cosmological term (and its
conjugate) as perturbation, for the non-unitary $(p,q)$
models. The three-point function of the cosmological operator can,
however, be computed in the latter approach and we
compare this result obtained in two different ways. We briefly discuss
the
$q\to\infty$ limit with fixed $p$, and get the characteristic
logarithmic singularity of the $c=1$ matter coupled to gravity.
We finally conclude with some speculative remarks.

The $(p,q)$ minimal matter, ($q>p$), is described by a
representation of the
Virasoro algebra with central charge
$$
  c = 1 - {6(q-p)^2\over qp}.
  \eqn\pqone
$$
States are built over the primaries $\phi_{r,s}$, ($1\le r\le q-1$ and
$1\le s\le p-1$), of conformal dimension
$$
  h_{r,s} = {(qs-pr)^2 - (q-p)^2\over 4qp}.
  \eqn\pqtwo
$$
When we couple this matter to gravity, the central charge of the
Liouville theory is given by $c_L = 26 - c = 1 + 6(q+p)^2/qp$.
Local operators in the combined theory have the form
\DDK\foot{We need not,
however, restrict the $r,s$ values inside the Kac table.}:
$\Phi_{r,s} = \phi_{r,s} e^{\lalrs\lphi}$, where,
$h_{r,s} + h(e^{\lalrs\lphi}) = 1$.
The dimension of the Liouville dressing operator follows from
this condition:
$$
  h(e^{\lalrs\lphi}) = {(q+p)^2 - (qs-pr)^2\over 4qp}.
  \eqn\pqsix
$$
In the Feigin-Fuks representation both matter and Liouville are
described
by free scalar fields with background charges $\qm = \sqrt{(1-c)/3} =
\sqrt{2/qp}(q-p)$ and $\ql = \sqrt{(25-c)/3} = \sqrt{2/qp}(q+p)$
respectively; and the local operators are vertex operators
$V^M_{(r,s)} = e^{i\malrs\mphi}$ and $V^L_{(r,s)} =
e^{\lalrs\lphi}$ respectively. The dimensions of the vertex
operators are given by
$$
  h(e^{i\malrs\mphi}) = \half\malrs(\malrs - \qm),\qquad\qquad
  h(e^{\lalrs\lphi}) = -\half\lalrs(\lalrs - \ql).
  \eqn\pqseven
$$
Solving from eqs.\pqsix\ and \pqseven, we find:
$$
  \lalrs{}^\pm = {(q+p)\pm(qs-pr)\over \sqrt{2qp}}.
  \eqn\pqeight
$$
The $(p,q)$ minimal model is in general non-unitary and has
operators with
both positive as well as negative dimensions. The cosmological
operator is
by definition the gravitationally dressed primary of minimum dimension
$\phi_{min}$. The conformal dimension $h_{min}$ of $\phi_{min}$ and the
corresponding Liouville momemta are given by:
$$
  h_{min} = {1-(q-p)^2\over 4qp},\qquad\qquad
  \alpha^L_{min}{}^\pm = {(q+p)\pm 1\over\sqrt{2qp}}.
  \eqn\pqnine
$$
The cosmological operator corresponds to the negative sign in
eq.\pqnine \ZEMICLASSICAL, and we denote it by $\Phi^-_{min}$:
$$
  \Phi^-_{min}(z,\bar z) = \phi_{min}e^{\alpha_{min}^{L-}\lphi}
  (z,\bar z).
  \eqn\pqten
$$
The corresponding coupling $\mu$ is called the cosmological
constant. For the unitary theories $(q=p+1)$, the identity operator
$\phi_{1,1}$ is the operator of minimum dimension $(h_{min}=0)$;
and for the
non-unitary $(2,q)$ series it is the operator $\phi_{(q-1)/2,1}$
$(=\phi_{(q+1)/2,1})$. For a general $(p,q)$ model, however, no general
expressions for $r$ and $s$ could be given, and one has to look for
$\phi_{min}$ case by case.

The Liouville action has the cosmological term $\mu\int d^2z
\Phi^-_{min}(z,\bar z)$, and in ref.\GOULIANLI\ it was used as the
screening charge in computing the correlation functions of the $(p,p+1)$
unitary minimal models coupled to gravity. Dotsenko \DOTSENKO\ has
proposed the following effective action for the Liouville theory:
$$
  S_L(\lphi) = {1\over 2\pi}\int
  d^2z\partial\lphi(z)\bar\partial\lphi(\bar z) +
  \mu\int d^2z\Phi^-_{min}(z,\bar z) +
  a\mu^\d\int d^2z\Phi^+_{min}(z,\bar z),
  \eqn\pqeleven
$$
where, $a$ is an arbitrary constant and $\d$ is the {\it gravitational
scaling dimension} of the operator $\Phi_{min}^+$ and is given by
$\d=\alpha^{L+}_{min}/\alpha^{L-}_{min}$. Calculation of correlation
functions can be done {\it perturbatively} and the cosmological term
$Q^-_{min}=\mu\int d^2z\Phi^-_{min}(z,\bar z)$ and its conjugate
$Q^+_{min}=a\mu^\d\int d^2z\Phi^+_{min}(z,\bar z)$ serve as screening
charges. The analysis of ref.\DOTSENKO\ is restricted to the case of
unitary theories where $\phi_{min}$ is the identity operator. Therefore
the presence of the screening charges $Q_{min}^\pm$ in the computation
of the correlation function in the Liouville sector does not modify the
matter sector contribution, which can be computed independently. For the
non-unitary theories however, $\phi_{min}$ is not the identity operator
and consequently in screening the Liouville sector by $Q_{min}^\pm$,
contribution from the matter sector gets modified. There are now some
extra $\phi_{min}$ whose effect has to be taken into account
in requiring
charge balance. It is easy to check that among the different possible
Feigin-Fuks representations for the various fields involved in the
computation of a general three-point function, there is not one which
screens both the Liouville and the matter sectors
simultaneously\foot{Correlators of the cosmological operators are
exceptions; and we shall come back to this point later.}. The remaining
option is to screen the (augmented) matter contribution by the usual
screening charges, {\it i.e.}, integrals of the dimension $(1,1)$ matter
operators \DOTSENKOFATEEV:
$$\eqalign{
  J^\pm(z,\bar z) &= e^{i\mb{}^\pm\mphi}\otimes \id_L(z,\bar z),\cr
  1 &= \half\mb(\mb-\qm).\cr}
  \eqn\pqtwelve
$$
There are now four types of screening currents
$\Phi_{min}^\pm(z,\bar z)$,
$J^\pm(z,\bar z)$. Whereas, the OPE of $\Phi_{min}^\pm$ (and $J^\pm$)
between themselves has the standard double pole singularity; the OPE
between $\Phi_{min}^+(z,\bar z)J^-(w,\bar w)$
(or $\Phi_{min}^-(z,\bar z) J^+(w,\bar w)$)
goes as a rational power of $(z-w)$ and the interchange of
their corresponding contours involves complicated phase factors. This
feature makes the generalization of the Dotsenko-Fateev integrals
\DOTSENKOFATEEV\ notoriously difficult. We do not pursue this approach
further.

We propose, instead, to work with the following effective action for the
Liouville theory:
$$\eqalign{
  S_L(\lphi) &= {1\over 2\pi}\int
  d^2z\partial\lphi(z)\bar\partial\lphi(\bar z) +
  A_-\mu^{\dnot^-}\int d^2z\id_M\otimes e^{\lb^-\lphi}(z,\bar z)\cr
  &\qquad\qquad\qquad+
  A_+\mu^{\dnot^+}\int d^2z\id_M\otimes e^{\lb^+\lphi}(z,\bar z),\cr}
  \eqn\pqthirteen
$$
where, $A_\pm$ are arbitrary constants and $\dnot^\pm$ are the
gravitational scaling dimensions of the $(1,1)$ Liouville operators
$e^{\lb^\pm\lphi}$:
$$
  \dnot^\pm = {\lb^\pm\over\alpha_{min}^{L-}}
  \eqn\pqfourteen
$$
and $\lb^\pm$ are such that the Liouville vertex operator is $(1,1)$:
$$
  -\half\lb(\lb-\ql) = 1.
  \eqn\pqfifteen
$$
The marginal operators $\Phi_0^\pm(z,\bar z) =
A_\pm\mu^{\dnot^\pm}e^{\lb^\pm\lphi}(z,\bar z)$ serve as the currents in
screening the Liouville contribution. We would like to emphasize that
for the non-unitary theories, these screening operators are not the
cosmological term \pqten\ and its conjugate,
but we have tuned their corresponding coupling constants
(which in general
are completely unconstrained) in a very particular $\mu$-dependent form.
The justification of this we can only provide {\it a posteriori}
in that it
produces the correct result. And finally note that since the matter
contribution to $\Phi_0^\pm$ is identity, we can (as in the unitary
theories), screen the matter sector independent of the Liouville with the
integrals of the standard screening currents $J_\pm$ \DOTSENKOFATEEV.

We now compute the three-point function $\l\Phi^-_{n',n}(z_1,\bar z_1)
\Phi^-_{s',s}(z_2,\bar z_2)\Phi^-_{m',m}(z_3,\bar z_3)\r$,
\foot{From now on
we shall restrict ourselves with operators $\Phi_{r,s}$ that have
Liouville momentum $\lalrs{}^-$ without any loss of generality\DOTSENKO.}
which by the
use of projective invariance reduces to the computation of the structure
constants:
$$\eqalign{
  \cc^{(m',m)}_{(n',n)(s',s)} &=
  \l\Phi^-_{n',n}(0)\Phi^-_{s',s}(1)\Phi^-_{m',m}(\infty)\r\cr
  &={\l\phi^-_{n',n}(0)\phi^-_{s',s}(1)\phi^-_{m',m}(\infty)\r}_M
  {\l e^{\laln\lphi}(0)e^{\lals\lphi}(1)e^{\lalm\lphi}(\infty)\r}_L\cr
  &\equiv C^{M(m',m)}_{(n',n)(s',s)}C^{L(m',m)}_{(n',n)(s',s)}.\cr}
  \eqn\pqsixteen
$$
$C^{M(m',m)}_{(n',n)(s',s)}$ can be computed as in ref.\DOTSENKOFATEEV\
using $N_+$ screening charges $\int d^2z J^+(z,\bar z)$ and $N_-$
screening charges $\int d^2z J^-(z,\bar z)$. The charge neutrality
condition
$\sum\malrs + N_+\mb^+ + N_-\mb^- = \qm$,
requires that
$$
  N_+ = {n+s+m-1\over 2},\qquad\qquad
  N_- = {n'+s'+m'-1\over 2}.
  \eqn\pqeighteen
$$
Using the integral formula (B.10) of ref.\DOTSENKOFATEEV, we get\foot{We
have omitted the $N_+! N_-!$ terms.}:
$$\eqalign{
  &C^{M(m',m)}_{(n',n)(s',s)} =\cr
  &\!\!
  {\pi^{N_-+N_+}\over\rho^{4N_-N_+}}
  {\left({\g {1-\rho '}\over\g {\rho '}}\right)}^{N_-}
  {\left({\g {1-\rho}\over\g {\rho}}\right)}^{N_+}
  \prod_{i=1}^{N_-}{\g {-N_++i\rho '}\over\g {1+N_+-i\rho '}}
  \prod_{j=1}^{N_+}{\g {j\rho}\over\g {1-j\rho}}\cr
  &\!\!\!\!\!\!
  \prod_{i=1}^{N_-}{\g {-N_+-n'\rho '+n+i\rho '}
  \g {-N_+-s'\rho '+s+i\rho '}
  \g {-N_+-m'\rho '+m+(N_-+1)\rho '-i\rho '}\over
  \g {1+N_++n'\rho '-n-i\rho '}\g {1+N_++s'\rho '-s-i\rho '}
  \g {1+N_++m'\rho '-m-(N_-+1)\rho '+i\rho '}}\cr
  &\!\!\!\!\!\!
  \prod_{j=1}^{N_+}{\g {n'-n\rho+j\rho}\g {s'-s\rho+j\rho}
  \g {m'-m\rho+j\rho}\over
  \g {1-n'+n\rho-j\rho}\g {1-s'+s\rho-j\rho}
  \g {1-m'+m\rho-j\rho}}\cr
  }
  \eqn\pqnineteen
$$
where, $\rho '=p/q=(\mb^-)^2/2$ and $\rho=q/p=(\mb^+)^2/2$.
The Liouville
structure constant $C^{L(m',m)}_{(n',n)(s',s)}$ requires $N_\pm^L$
screening charges $\int d^2z\Phi_0^\pm(z,\bar z)$, where the charge
neutrality condition in the Liouville sector requires that:
$$\eqalign
  N_+^L = {n+s+m-1\over 2} = N_+,\qquad\qquad
  N_-^L = -{n'+s'+m'+1\over 2} = -(N_-+1).
  \eqn\pqtwenty
$$
Using the integral formula and the analytic continuation
for products of a
negative number of terms as given in \DOTSENKO, we get:
$$\eqalign{
  &C^{L(m',m)}_{(n',n)(s',s)} =\cr
  &\!
  A_-^{-(N_-+1)}A_+^{N_+}\mu^{-(N_-+1)\dnot^-+N_+\dnot^+}
  \pi^{-(N_-+1)+N_+}\rho^{4(N_-+1)N_+}
  {\left({\g {-\rho '}\over\g {1+\rho '}}\right)}^{N_-+1}
  {\left({\g {1+\rho}\over\g {-\rho}}\right)}^{N_+}\cr
  &\!\!\!\!
  \prod_{i=0}^{N_-}{\g {1+N_+-i\rho '}\over\g {-N_++i\rho '}}
  \prod_{j=1}^{N_+}{\g {-j\rho}\over\g {1+j\rho}}\cr
  &\!\!\!\!
  \prod_{i=0}^{N_-}{\g {1+N_++n'\rho '-n-i\rho '}
  \g {1+N_++s'\rho '-s-i\rho '}
  \g {1+N_++m'\rho '-m-N_-\rho '+i\rho '}\over
  \g {-N_+-n'\rho '+n+i\rho '}\g {-N_+-s'\rho '+s+i\rho '}
  \g {-N_+-m'\rho '+m+N_-\rho '-i\rho '}}\cr
  &\!\!\!\!
  \prod_{j=1}^{N_+}{\g {-n'+n\rho-j\rho}\g {-s'+s\rho-j\rho}
  \g {-m'+m\rho-j\rho}\over
  \g {1+n'-n\rho+j\rho}\g {1+s'-s\rho+j\rho}
  \g {1+m'-m\rho+j\rho}}\cr
  }
  \eqn\pqtwentyone
$$
Multiplying expressions \pqnineteen\ and \pqtwentyone\ (using
eq.\pqsixteen), we get the unnormalized structure constants:
$$\eqalign{
  \cc^{(m',m)}_{(n',n)(s',s)} &= (-1)^{N_-+N_++1}
  A_-^{-(N_-+1)}A_+^{N_+}
  {\pi^{2N_+-1}\over\g 0}\rho^{2(N_--N_++1)}
  {\left({\g {1-\rho '}\over\g {\rho '}}\right)}^{2N_-+1}\cr
  &\qquad\times
  \mu^{-(N_-+1)\dnot^-+N_+\dnot^+}\;
  {\g {1+n'\rho '-n}\g {1+s'\rho '-s}\g {1+m'\rho '-m}\over
  \g {-n'\rho '+n}\g {-s'\rho '+s}\g {-m'\rho '+m}}.\cr
  }
  \eqn\pqtwentytwo
$$
Note that this expression agrees with that of ref.\DOTSENKO\ in the case
$q=p+1$. Now to compare this general result with the matrix model result
we proceed to calculate the normalization independent structure
constants of the $(p,q)$ minimal matter coupled to gravity.
For that we need to know
the two point functions of the operators involved and the partition
function. The two point function is given by:
$$
  -{d\over d\mu}\l\Phi_{n',n}\Phi_{n',n}\r = A_-\dnot^-\mu^{\dnot^--1}
  \l\Phi_{n',n}\Phi_{n',n}\Phi_{1,1}\r.
  \eqn\pqtwentythree
$$
Using eq.\pqtwentytwo and integrating the above expression we derive the
two point function as:
$$\eqalign{
  \l\Phi_{n',n}\Phi_{n',n}\r &=
  A_-\dnot^-{\mu^{\dnot^-}\over n'\dnot^--n\dnot^+}
  \l\Phi_{n',n}\Phi_{n',n}\Phi_{1,1}\r\cr
  &= {(-1)^{n'+n-1}A_-^{-n'}A_+^n\pi^{2n-1}\over\rho^{2(n'-n+1)}}\;
  {\dnot^-\mu^{-n'\dnot^-+n\dnot^+}\over n'\dnot^--n\dnot^+}
  {\left({\g {1-\rho '}\over\g {\rho '}}\right)}^{2n'}
  \left({\g {1+n'\rho '-n}\over\g {-n'\rho '+n}}\right)^2.\cr}
  \eqn\pqtwentyfour
$$
Similarly the partition function is given by
$$
  -{d^3\over d\mu^3}Z = \left(A_-\dnot^-\mu^{\dnot^--1}\right)^3
  \l\Phi_{1,1}\Phi_{1,1}\Phi_{1,1}\r
  \eqn\pqtwentyfive
$$
which after integration gives the following expression:
$$\eqalign{
  Z &= -\;{\left(A_-\dnot^-\mu^{\dnot^-}\right)^3\over
  (\dnot^-+\dnot^+-2)(\dnot^-+\dnot^+-1)(\dnot^-+\dnot^+)}
  \l\Phi_{1,1}\Phi_{1,1}\Phi_{1,1}\r\cr
  &= A_-A_+
  {(\dnot^-)^3\mu^{\dnot^-+\dnot^+}\over
  (\dnot^-+\dnot^+-2)(\dnot^-+\dnot^+-1)(\dnot^-+\dnot^+)}
  \;{\pi\over\g 0}\rho^2.\cr
  &= A_-A_+{\pi\over\g 0}{2q^2p\over (q+p)(q+p+1)}\;
  \mu^{2(q+p)/(q+p+1)}\cr
  }
  \eqn\pqtwentysix
$$
Substituting from the expressions \pqtwentytwo, \pqtwentyfour\ and
\pqtwentysix, we get the expression for the normalized three-point
function:
$$\eqalign{
  \cd^{(m',m)}_{(n',n)(s',s)} &\equiv
  {{\l\Phi_{n',n}\Phi_{s',s}\Phi_{m',m}\r}^2 Z\over
  \l\Phi_{n',n}\Phi_{n',n}\r \l\Phi_{s',s}\Phi_{s',s}\r
  \l\Phi_{m',m}\Phi_{m',m}\r}\cr
  &= -\;{(n'\dnot^--n\dnot^+)(s'\dnot^--s\dnot^+)(m'\dnot^--m\dnot^+)
  \over (\dnot^-+\dnot^+-2)(\dnot^-+\dnot^+-1)(\dnot^-+\dnot^+)}\cr
  &= -\;{2(n'p-nq)(s'p-sq)(m'p-mq)\over (q+p)(q+p+1)}.\cr}
  \eqn\pqtwentyseven
$$
This result for the general $(p,q)$ minimal models is in exact agreement
with the conjecture of
Imbimbo and Mukhi \PRIVATE.\foot{We thank S.~Mukhi
for pointing this to us.}

For the special case of the unitary theories $(q=p+1)$, the marginal
deformation we have considered (eq.\pqthirteen) coincides with the
cosmological terms (as identity is the operator with the least dimension
for such theories). In these theories $\dnot^-=1$ and $\dnot^+=\delta$
(from eq.\pqfourteen); and we recover the result of ref.\DOTSENKO:
$$
  \cd^{(m',m)}_{(n',n)(s',s)}{\Big|}_{(p,p+1)} =
  {(n'-n\d)(s'-s\d)(m'-m\d)\over\d(1+\d)(1-\d)},
  \eqn\pqtwentyeight
$$
which agrees with the matrix model results \DIFRKUTMATRIX.

On the other hand, for the $(2,2k-1)$ non-unitary minimal matter, the
basic operators are of the form $\Phi_{r,1}$, with $r=k-1$ being the
operator of least dimension. For these theories $\dnot^-=2/k$ and
$\dnot^+=2-1/k$. Substituting in eq.\pqtwentyseven, we get the
normalized three-point function:
$$
  \cd^{(m,1)}_{(n,1)(s,1)}{\Big|}_{(2,2k-1)}
  =-{(2n-2k+1)(2s-2k+1)(2m-2k+1)\over(2k+1)(k+1)}.
  \eqn\pqtwentynine
$$
The above result can be compared with the following
one-matrix model results
of ref.\GROSSMIGDAL\ (notation as in \GROSSMIGDAL):
$$\eqalign{
  F &= {k^2\over (k+1)(2k+1)}\; t^{2+{1\over k}}\cr
  \l\co_{l_1}\r &= -{k\over(l_1+1)(l_1+k+1)}\; t^{1+{l_1+1\over k}}\cr
  \l\co_{l_1}\co_{l_2}\r &= -{1\over(l_1+l_2+1)}\;
  t^{{l_1+l_2+1\over k}}\cr
  \l\co_{l_1}\co_{l_2}\co_{l_3}\r &= -{1\over k}\;
  t^{-1+{l_1+l_2+l_3+1\over k}}\cr}
  \eqn\pqthirty
$$
where, $l_r$ labels the operators and $t$ is the cosmological constant,
which is the coupling corresponding to the operator $\co_0$.
Comparison of the matrix model free energy $F(t)$ with our
partition function shows that
apart from normalization (which are obviously different in the two
different formalisms), the matrix model cosmological constant
$t$ is same as $\mu$. We can also work out the normalization
independent three-point function from eq.\pqthirty:
$$
  {{\l\co_{l_1}\co_{l_2}\co_{l_3}\r}^2 F\over \l\co_{l_1}\co_{l_1}\r
  \l\co_{l_2}\co_{l_2}\r\l\co_{l_3}\co_{l_3}\r} =
  -\;{(2l_1+1)(2l_2+1)(2l_3+1)\over (k+1)(2k+1)}.
  \eqn\pqthirtyone
$$
The above agrees with our continuum result eq.\pqtwentynine\ with the
identification:
$$
  l_r = k-1-r:\qquad
  \co_{l_r} \sim \Phi_{k-1-r,1}.
  \eqn\pqthirtytwo
$$
This identification is natural if we remember that $r=k-1$,
($l_{k-1}=0$), is the minimum dimension operator in the
$(2,2k-1)$ model, and it is natural
to label the fields starting from the cosmological operator.

Let us now comment on the computation of the correlators of the
cosmological operator $\Phi_{min}^-$ using the action \pqeleven. In
this approach $Q_{min}^\pm\sim\int d^2z\Phi_{min}^\pm (z,\bar z)$
serve as
the screening charges, and fortunately here it is possible to screen the
matter and the Liouville parts simultaneously using these charges. The
$n$-point correlator of $\Phi_{min}^-$ for example requires one
$Q_{min}^+$ and $-(n-1)$ number of $Q_{min}^-$ for screening.
It is straightforward to work out
the three-point function:
$$
  \l\Phi_{min}^-(0)\Phi_{min}^-(1)\Phi_{min}^-(\infty)\r =
  {1\over\pi\g 0}{(1+q)^8(1-q)^2\over q^{12}}\;\mu^{-(q+p-3)/(q+p-1)}
  \eqn\pqthirtyfive
$$
which upon integrating three times with respect to $\mu$ gives the
partition function:
$$
  Z=-{1\over\pi\g 0}{(1+q)^8(1-q)^2\over q^{12}}
  {(q+p-1)^3\over 4(q+p)(q+p+1)}\;
  \mu^{2(q+p)/(q+p-1)}.
  \eqn\pqthirtysix
$$
Comparing this expression with \pqtwentysix, we see that they differ by
normalization. But in the case of \pqtwentysix, we have the freedom to
choose the arbitrary constants $A_\pm$ so that the partition function we
have calculated using the action \pqthirteen\ agrees with that
calculated
by integrating the three-point function of the cosmological operator. We
could also view this as a check for our ansatz \pqthirteen.
Unfortunately, however, it is not possible to compute arbitrary
three-point functions in
the alternative approach using \pqeleven\ due
to technical difficulties discussed in the beginning.

Finally we consider the limit $q\to\infty$ with $p$ fixed. Although the
central charge $c$ of the matter part goes to $-\infty$ in this limit,
the effective central charge $c_{eff}=c-24h_{min}=1-6/{qp}$,
which measures the degrees of freedom of the theory, tends to the
limiting value $1$. We
would therefore expect to get the results related to that of
$c=1$ matter coupled to gravity \MATRIXTWO.
\foot{A space-time
interpretation of the $(p,q)$ minimal matter coupled to gravity was
developed in \IMBIMUKHI, where the limit $q\to\infty$ was
discussed from a string field theoretic point of view.}
We may also recall that the one-loop partition function contains
information about the operator content of a theory and this quantity for
the $(p,q)$ model can be written as the difference of the partition
functions of scalar fields taking values on circles of radii
$R_1=\sqrt{2pq}$ and $R_2=\sqrt{2p/q}$ respectively \DIFRANSALZ. In the
limit $q\to\infty$, the second radius $R_2\to 0$, and the first radius
$R_1$ diverges with the scalar field taking all real values.
The partition function of a free scalar field, (taking values on
the real line), coupled to gravity has been calculated
\KAZAMIG\MATRIXTWO\
and found to have a characteristic logarithmic singularity:
$$
  Z_{c=1}\sim\mu^2\ln\mu.
  \eqn\pqthirtythree
$$
On the other hand, if we take the limit $q\to\infty$
with fixed $p$, we get:
$$\eqalign{
  Z\; &\;{\buildrel {q\to\infty}\over\longrightarrow}\;
  A_-A_+{\pi\over\g 0}\; 4\left({p\over q}\right)\;\mu^2\ln\mu\cr
  \cd_{(n',n)(s',s)(m',m)}
  &\;{\buildrel {q\to\infty}\over\longrightarrow}\;
  4\left(n-{n'p\over q}\right)\left(s-{s'p\over q}\right)
  \left(m-{m'p\over q}\right)\;\ln\mu\cr}
  \eqn\pqthirtyfour
$$
The $p/q$ factor in the front of $Z$ seems to be an artifact of the
calculation as it drops out in the normalized three-point function. The
normalized correlation function also agrees with three-point function of
tachyon operators \SAKAITANII\ if the momentum of the tachyon $k_{r,s}$
is identified with ${2\over \sqrt{\alpha'}}\left(s-r{p\over q}\right)$.
No analog of the momentum conservation condition however follows
from this limiting procedure.

We have computed the tree-level three-point correlation functions
of local operators
in the general $(p,q)$ non-unitary minimal models coupled to gravity. We
have used the free Liouville action perturbed by the marginal operators
which are integrals of gravitationally dressed matter identity.
These operators are not the cosmological operators, but their
corresponding couplings are tuned to be functions of the cosmological
constant $\mu$. This perturbation is much easier to work with
rather than
the perturbation by the cosmological operators, which is a direct
generalization of the action suggested by Dotsenko\DOTSENKO.
The reason is that in the case of non-unitary theories the
matter operator
of least dimension is not the identity and this modifies the matter
contribution, which needs further screening. The evaluation of the
corresponding Dotsenko-Fateev integrals become extremely complicated to.
However, the correlators of the cosmological operator can be
computed in both the approaches and we have compared the two results. We
have also shown that our expression for the $(2,2k-1)$ minimal series
agrees with the one-matrix model results \GROSSMIGDAL. For the unitary
series $q=p+1$, our action coincides with that of Dotsenko \DOTSENKO.

The justification of our proposed action could be given
{\it a posteriori} that it reproduces the correct answer.
However, since the perturbation we
consider is not by the cosmological operators, it is tempting to
conjecture that any marginal perturbation of the free Liouville action,
with the coupling properly adjusted,
would reproduce the correct tree-level three-point correlation
functions. Such a possibility has also been mentioned by Dotsenko
(see the second reference in \DOTSENKO).
The perturbation we have considered is, however, technically
the simplest
to work with (as it lets one screen the Liouville and the matter
independently). The most natural Coulomb gas formalism for the minimal
matter coupled to gravity should presumably follow from some BRST-like
symmetry analogous to that of ref.\FELDER\ for the case of the minimal
models.

\noindent Acknowledgement: It is a pleasure to thank V.~Dotsenko,
S.~Mukhi, A.~Sen and A.~M.~Sengupta for many interesting discussions.

\refout
\bye